\documentclass[useAMS,usenatbib,usegraphicx]{mn2e}

\usepackage{amssymb}
\usepackage{amsmath}
\usepackage{url}

\def\fermi{{\it Fermi\/}}

\title[Search for dark matter in HI clouds]{Search for dark 
matter in compact hydrogen clouds}
\author[N. Mirabal]{N. Mirabal$^{1,2}$\thanks{E-mail:
mirabal@gae.ucm.es}\\
$^{1}$Ram\'on y Cajal Fellow\\
$^{2}$ Dpto. de F\'isica At\'omica,
Molecular y Nuclear, Universidad Complutense de
Madrid, Spain\\
}

\begin{document}

\date{}

\pagerange{\pageref{firstpage}--\pageref{lastpage}} \pubyear{2013}

\maketitle

\label{firstpage}

\begin{abstract}
The recently published GALFA-HI Compact Cloud Catalogue lists 20 
neutral hydrogen clouds that might pinpoint previously undiscovered  
high-latitude dwarf galaxies. Detection of  
an associated gamma-ray dark matter signal  
could provide a route to distinguish unambiguously 
between truly dark matter dominated systems that
have accumulated neutral hydrogen but have not successfully ignited 
star formation and pure gaseous structures devoid of dark matter. We 
use 4.3 years of {\it Fermi} observations to derive gamma-ray 
flux upper limits in the 1--300 GeV energy range
for the sample. Limits on gamma rays from pair 
 annihilation of dark matter are also presented depending on 
the yet unknown
astrophysical factors. 
\end{abstract}


\section{Introduction}
The number of dark matter subhalos surrounding the Milky Way 
has puzzled dark matter aficionados 
for over more than a decade. 
While cosmic large-scale structures are well described by the Lambda Cold
Dark Matter ($\Lambda$CDM) cosmological model, a 
number of discrepancies exist between the standard theory of galaxy formation 
and observations of substructures at smaller scales \citep{frenk}. 
In particular, numerical $\Lambda$CDM simulations consistently predict that 
galaxies must be surrounded by a huge population 
of subhalos \citep{klypin,moore}. 
Intuitively, subhalos would encompass anything from the largest satellites
({\it e.g.} dwarf galaxies) 
to substructures with masses around $10^{-4} M_{\odot}$ \citep{loeb}. 
Unfortunately, completing a survey of the smallest gravitationally bound 
systems is not straightforward \citep{ando2}. At faint flux levels, 
it becomes ever more difficult to recognise sparsely populated stellar  
systems. Important progress has been made recently resulting from newly 
discovered ultra-faint dwarf galaxies, which have nearly doubled
the number of known dwarfs \citep{willman,belokurov}. 
Yet the dwarf counts in 
the Milky Way 
remains far to small compared to the number of subhalos 
predicted by simulations.

Interestingly, 
the discovery of compact hydrogen clouds potentially located in the
Galactic halo has revived the possibility of a larger population of   
galaxy candidates that could be traced by neutral hydrogen
\citep{lockman,ryan}. This point was made even
earlier by \citet{klypin} who
recognised that the much broader 
and physically distinct 
set of enigmatic 
HI structures commonly referred to as high-velocity clouds (HVCs) 
could represent the missing dark matter subhalo population 
\citep{oort,bregman,braun,blitz}.  Although intriguing, to date, 
a firm HI cloud-subhalo link has not
been established \citep{quilis}.

Prompted by the recent release of the
GALFA HI Compact Cloud Catalogue \citep{saul}, we revisit the possibility 
that some compact hydrogen clouds 
can be used as a proxy for missing dark matter subhalos.
There are inherent problems when trying to test this possibility, especially
the lack of distance and mass information for such systems.
Another major hurdle is the  
absence of associated stellar populations in most compact hydrogen
clouds. In fact, it is even  possible that many of these systems 
never formed stars \citep{ricotti}. \citet{lewis} 
proposed exploiting
``pixel gravitational lensing'' as a way to map the 
dark matter content in hydrogen clouds. 
However,
the shortage of properly aligned nearby background galaxies
prevents a generalised application to a large  
sample.

 \begin{figure*}
\hfil
\includegraphics[width=5.0in]{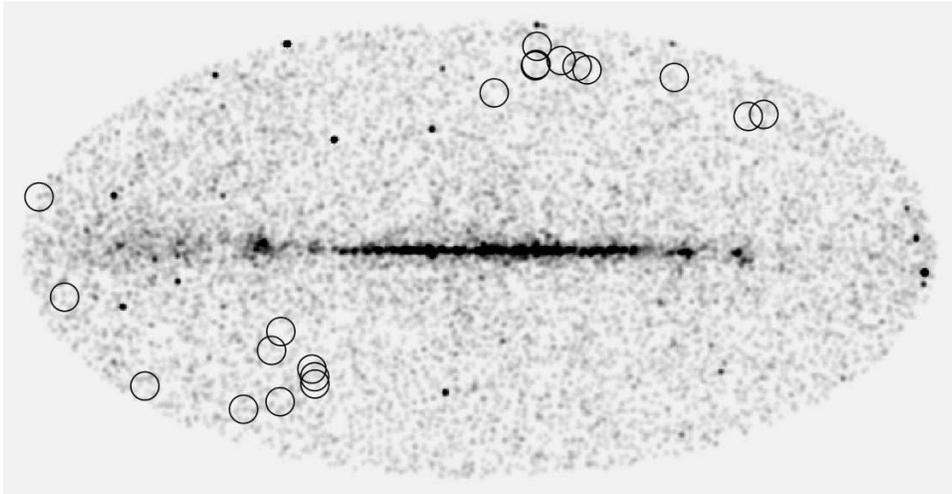}
\hfil
\caption{Aitoff projection
of the {\it Fermi} all-sky LAT map, showing the
locations of the 20 galaxy candidates in our study.
Galactic coordinates.}
\label{figure1}
\end{figure*}

For this reason, we explore a different approach to determine 
their dark matter content.
If compact hydrogen clouds are dark matter dominated systems, 
nearby dense 
objects could potentially produce a detectable dark matter annihilation signal 
in gamma rays \citep{berg,baltz}. 
In that vein, \citet{flix} looked for spatial coincidences 
between unidentified EGRET sources and HVCs.  With the vast improvements in
gamma-ray 
sensitivity and angular resolution afforded by {\it Fermi} \citep{atwood},  
we can search for annihilating dark matter 
in GALFA-HI compact hydrogen clouds directly.  This study 
should serve not only to investigate a possible tracer of 
gas-bearing
dark matter seeds that never formed stars, 
but also to set a general upper bound on the annihilation cross section 
$\langle\sigma v\rangle_{\chi}$ of a hypothesised 
weakly interacting massive particle 
(WIMP). 
Searches for dark matter signals using
{\it Fermi} have been conducted unsuccessfully in a wide variety 
of astrophysical
systems  \citep{buckley,ackermann,ando,mirabal}. 
This work simply extends the hunt to a new sample.

The paper is structured as follows. In Section \ref{sec2} we
explain the selection of high-latitude 
galaxy candidates from the GALFA-HI Compact Cloud
Catalogue, as well as the {\it Fermi} LAT analysis. 
In Section \ref{sec3} we derive gamma-ray flux upper limits for Segue 1 and set
dark matter annihilation constraints for the galaxy candidates.
Finally, we briefly summarise
our interpretation and possible future directions in Section \ref{sec4}.

\section{Galaxy candidates and {\it Fermi}  LAT analysis}
\label{sec2}
The GALFA-HI Compact Cloud Catalogue \citep{saul} is generated from the 
Galactic Arecibo $L$-Band Feed Array HI (GALFA-HI) Survey Data Release 
One \citep{peek}.
At completion, GALFA-HI will cover 13,000 deg$^{2}$ of the sky
in the 1420 MHz 
hyperfine transition of hydrogen  between $V_{LSR} = \pm 650$ km s$^{-1}$. 
Using a novel cloud detection algorithm, \citet{saul}
identified a total of 1964 compact ($< 20^{\prime}$) hydrogen clouds in the 
initial 7520 deg$^{2}$ Data Release One. The catalogue breaks down the 
clouds into a scheme that includes high-velocity clouds (HVCs), galaxy 
candidates, cold low-velocity clouds (CLVC), warm low-velocity clouds,
and warm positive low-velocity clouds in the third Galactic quadrant.  
For our purposes, we are only concerned with the 27 possible 
galaxy candidates that might form
the core sample of potentially undiscovered dark matter subhalos. 
In order to guard against possible gaseous disk interlopers 
that may have been pushed into the halo 
by stellar feedback \citep{ford}, we also 
prune galaxy candidates at $|b|\leq 10^\circ$. 
Finally, we are left with a subset of 20 high-latitude galaxy candidates.
In Fig. \ref{figure1}, we show the distribution of  
these systems on the sky.  

\begin{table}
\caption{Gamma-ray flux upper limits at 95\% confidence level.}
\begin{tabular}{l c c c}
\hline
Galaxy Candidate & $l$ ($^\circ$) & $b$ ($^\circ$) & $F_{\rm lim}$ (1--300 GeV) \\
                    &    &   &  (ph cm$^{-2}$ s$^{-1}$)\\
\hline
003.7+10.8+236 & 108.53 & -51.02  & 6.4 $\times 10^{-11}$\\
019.8+11.1+617 & 133.84 & -51.16  & 5.5 $\times 10^{-11}$\\
044.7+13.6+528 & 164.15 & -38.83  & 1.1 $\times 10^{-10}$\\
063.7+33.3+447 & 164.58 & -12.64  & 2.8 $\times 10^{-10}$\\
100.0+36.7+417 & 178.44 & 13.67  & 1.1 $\times 10^{-10}$\\
143.7+12.9+223 & 220.08 & 41.96  & 6.6 $\times 10^{-11}$\\
147.0+07.1+525 & 228.97 & 42.22   & 4.1 $\times 10^{-11}$\\
162.1+12.5+434 & 233.73 & 57.73   & 6.3 $\times 10^{-11}$\\
183.0+04.4--112 & 278.82 & 65.41  & 8.0 $\times 10^{-11}$\\
184.8+05.7--092 & 281.76 & 67.21  & 1.1 $\times 10^{-10}$\\
187.5+08.0+473 & 287.03 & 70.18  & 4.5 $\times 10^{-11}$\\
188.9+14.5+387 & 285.67 & 76.84  & 6.0 $\times 10^{-11}$\\
195.9+06.9--100 & 311.64 & 69.59  & 4.0 $\times 10^{-11}$\\
196.6+06.5--105 & 313.31 & 69.06  & 4.1 $\times 10^{-11}$\\
215.9+04.6+205 & 351.18 & 58.53  & 7.3 $\times 10^{-11}$\\
331.8+21.0+303 & 79.13  & -27.58  & 5.9 $\times 10^{-11}$\\
339.0+09.0--237 & 76.00 & -41.18    & 9.1 $\times 10^{-11}$\\
341.7+07.7--234 & 77.58 & -43.95  & 3.9 $\times 10^{-11}$\\
342.1+20.6+208 & 87.76 & -33.78  & 7.3 $\times 10^{-11}$\\
345.0+07.0--245 & 80.58 & -46.48  & 4.1 $\times 10^{-11}$\\
\hline
\end{tabular}
\label{table1}
\end{table}

In order to explore the gamma-ray emission, we use the publicly available 
dataset acquired by the Large Area Telescope (LAT) instrument
on board the \fermi\ Gamma-ray Space Telescope \citep{atwood}. 
The LAT is a pair-conversion gamma-ray detector sensitive 
to photon energies from 20 MeV to 300 GeV.
We retrieve all photons of `source' class (\texttt{evclass=2}) 
within
a 10$^{\circ}$ circular region centred at the 
position of each galaxy candidate. The data analysed here were 
collected between 2008 August 4 and 2012 November 20 (approximately 4.3 years
of data). 
Good time intervals were processed using the available 
\texttt{v9r27p1} {\it Fermi} Science Tools with 
the standard
\texttt{P7SOURCE\_V6} instrument response function.
Throughout, we  
apply a maximum zenith angle cut of  100$^{\circ}$.
We further filter the data using the \texttt{gtmktime} filter expression
recommended by the LAT team, namely ``{(DATA\_QUAL==1) \&\& (LAT\_CONFIG==1)
\&\& ABS(ROCK\_ANGLE)$<$52}''
The final analysis for each region includes all the point sources listed 
in the 2FGL catalogue \citep{nolan}, the current
Galactic
diffuse emission model \texttt{gal\_2yearp7v6\_v0.fits},
and the extragalactic isotropic model \texttt{iso\_p7v6source.txt}.

\begin{figure}
\hfil
\includegraphics[width=3.1in,angle=0.]{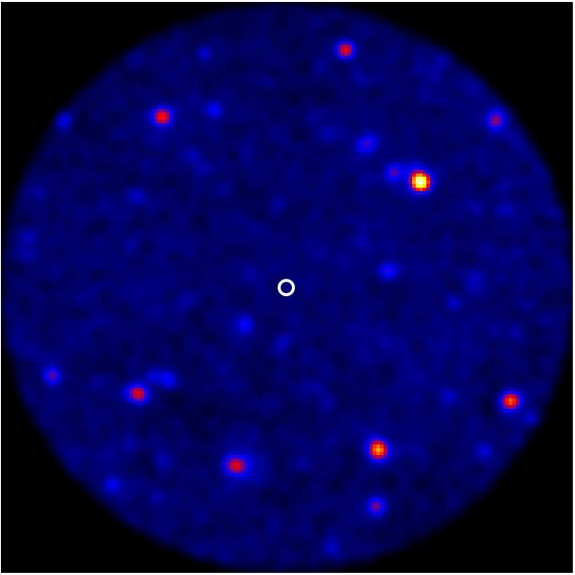}
\hfil
\caption{Smoothed {\it Fermi}-LAT (1--300 GeV) count map of 
galaxy candidate 143.7+12.9+223. 
The small white circle marks 
the centre of the Region of Interest (ROI). The map 
corresponds to a 20$^{\circ}$ circular region.}
\label{figure2}
\end{figure}

The resulting dataset is analysed with a binned likelihood method 
using the \texttt{gtlike} tool in the 
standard {\it Fermi} Science Tools\footnote{\tiny http://fermi.gsfc.nasa.gov/ssc/data/analysis/scitools/binned\_likelihood\_tutorial.html}.
For each position, we create a count map made up of 
30 logarithmically uniform energy bins using \texttt{gtbin}.   
We next construct a binned exposure map with  \texttt{gtexpcube2}, and  
a model source and diffuse 
count map with \texttt{gtsrcMaps}. 
Fig. \ref{figure2} shows a typical {\it Fermi} count map from
the sample. In the absence of emission, flux upper limits are then derived
using the implementation of 
\texttt{LATAnalysisScripts}\footnote{\tiny http://fermi.gsfc.nasa.gov/ssc/data/analysis/scitools/LATAnalysisScripts.html}. This set of Python 
libraries unifies the \texttt{pyLikelihood} module 
included in the standard {\it Fermi} Science tools. Upper limits
are computed with \texttt{calcUpper} assuming a power law spectrum
of high-energy emission $E^{-2}$
within a radius of 10$^{\circ}$. We
restrict our analysis to photons  in the 1--300 GeV energy range. 
The {\it Fermi} LAT point spread function (PSF) is
typically $0.8\,(E/1$GeV$)^{-0.8}$\,deg, which in our selected energy
range restricts the photons to less than 1$^{\circ}$ around each
location \citep{geringer}. 
Table \ref{table1} summarises the
95\% confidence level upper limits. Since the {\it Fermi} exposure is rather 
uniform over the sky the upper flux limits are fairly similar,  
except for a handful of locations with higher diffuse 
background emission or neighbouring bright
gamma-ray sources.

\section{Dark matter constraints}
\label{sec3}
If compact hydrogen clouds are highly dark matter dominated objects their
gamma-ray emission should be well characterised  
by a  differential spectrum 
that can be written as

\begin{equation}\label{eq1}
\frac{d\Phi}{dE}(E,\Delta\Omega) = \frac{1}{4\pi}
\frac{\langle\sigma v\rangle_{\chi}}{2 m^2_\chi}
\frac{dN_{\gamma}}{dE} \times J(\Delta\Omega),
\end{equation}

\noindent
where $\langle\sigma v\rangle_{\chi}$ is the
thermally averaged annihilation cross section, 
$m_{\chi}$ is the dark matter particle mass, and
$\frac{dN_{\gamma}}{dE}$ is the photon spectrum of
annihilation products \citep{abdo}.
The second term, or the so-called astrophysical factor $J(\Delta\Omega)$, 
corresponds to the
integration of the dark matter density squared $\rho^2(l,\Omega)$  along the 
line of sight $l$ of the compact cloud,
over a solid angle $\Delta\Omega$, so that
\begin{equation}\label{eq2}
J(\Delta\Omega)=
\int_{\Delta\Omega}\int \rho^2(l,\Omega)\,dl\,d\Omega.
\end{equation}

In principle, we expect compact clouds to be virtually free of
gamma-ray emission from embedded diffuse and 
individual point sources. As a result, it is
reasonable to expect that  
the differential spectrum from dark
matter annihilation should dominate any 
gamma-ray signal in the direction of observation to these systems. 

Formally, a robust computation of an
upper limit on the annihilation cross section
$\langle\sigma v\rangle_{\chi}$  
requires some knowledge of the
astrophysical factor $J(\Delta\Omega)$ of the system under consideration.
Given a lack of direct observational constraints on the astrophysical factors
of these galaxy candidates $J_{gc}$, 
we are only able to estimate $\langle\sigma v\rangle_{\chi}$ bounds from 
our sample 
by tying them to potentially similar systems in the Milky Way. 
Under the Ansatz that they are 
strongly dark matter dominated, the newly discovered
ultra-faint dwarf galaxies would appear to be the 
closest relatives to compact hydrogen clouds \citep{strigari}. 
As shown by \citet{simon2}, Segue 1 
represents the darkest of these systems with a very high
mass-to-light ratio  ($\sim$ 3400 M$_{\odot}$/L$_{\odot}$)
and dark matter density $2.5^{+4.1}_{-1.9}$ M$_{\odot}$ pc$^{-3}$.
It also boasts the largest astrophysical factor for known dwarfs
$J_{Segue 1} = 10^{19\pm 0.6}$ GeV$^{2}$ cm$^{-5}$
\citep{essig}.

Hereafter, we derive upper limits on $\langle\sigma v\rangle_{\chi}$
relative to the bounds already imposed for Segue 1
using {\it Fermi} measurements \citep{abdo,essig,scott,geringer}. 
Accordingly, we repeat the previous {\it Fermi} LAT analysis 
now centred on Segue 1 $(\ell, b)=(220.5^{\circ},50.4{^\circ})$.  
The corresponding gamma-ray upper limit for Segue 1 is $F_{\rm lim~Segue 1}$ 
(1--300 GeV) = 3.5 $\times 10^{-11}$ ph cm$^{-2}$ s$^{-1}$. In order to
turn our flux upper limits into a  bound on $\langle\sigma v\rangle_{\chi}$, 
we need to assume a specific annihilation channel. 
We adopt the strictest possible limit on 
dark matter annihilation into bottom
quarks $b\bar{b}$ for Segue 1 based on 3 years of {\it Fermi} data,
which translates into 
$\langle\sigma v\rangle_{\chi} 
\lesssim 3 \times 10^{-26}$
cm$^3$ ~ s$^{-1}$ for particle
masses $m_{\chi} \lesssim 40$ GeV \citep{geringer}. Assuming the annihilation
to be purely into $b\bar{b}$ and an
average flux for the galaxy candidates  
$\langle F_{lim}(gc) \rangle = 7.7 \times 10^{-11}$ ph cm$^{-2}$ s$^{-1}$,
 we can approximate $\langle \sigma v
 \rangle_{\chi}$ from these systems as 

\begin{equation}
\langle \sigma v
 \rangle_{\chi} \lesssim 7 \times 10^{-26} \frac{J_{Segue 1}}{J_{gc}} cm^3 ~ s^{-1},
\end{equation}

\noindent
for $m_{\chi} \lesssim 40$ GeV  WIMP masses annihilating to $b\bar{b}$, 
depending on the yet unknown
$J_{gc}.$

\section{Discussion and future prospects}
\label{sec4}
We have reported gamma-ray flux upper limits for a 
subset of galaxy candidates from the GALFA-HI 
Compact Cloud Catalogue. 
It is difficult to imagine any of these systems beating the
astrophysical factors of known dwarf galaxies that spans the gamut 
from $4\times 10^{17}$ GeV$^{2}$ cm$^{-5}$
for Carina  to $1.3\times 10^{19}$ GeV$^{2}$ cm$^{-5}$ for
the ultra-faint dwarf galaxy Segue 1 \citep{essig}.
Without mass and distance constraints, it will be difficult 
to place compact clouds
in the context of other measurements. 
It is generally assumed that in order to be
self gravitating, compact clouds
could pack masses as low as a few
M$_{\odot}$  to greater than
$\sim 10^{6}$ M$_{\odot}$ for outer systems \citep{giovanelli,saul}.
Our naive expectation is that  
$J_{Segue 1}/J_{gc} \gg 1$ should hold in most
cases. 

As an illustration, the average compact cloud 
properties indicate a mass of $\sim 2\times 10^{4}$ M$_{\odot}$ and a physical
size of $\sim 100$ pc at distance of 100 kpc \citep{saul}. Since the 
astrophysical factor is proportional to the density squared,  
we can roughly approximate $J_{Segue 1}/J_{gc} \approx 10^{3}$ for 
$M_{Segue 1} \sim 6\times 10^{5}$ M$_{\odot}$ \citep{simon2}. 
To be sure, arguments 
consistent with an interpretation that 
Segue 1 is a tidally disrupting star cluster
contaminated by the Sagittarius stream should be definitively ruled out 
\citep{niederste}. 
For the entire family of ultra-faint galaxies, 
it is also critical to derive more reliable estimates 
of the $J$-factor \citep{walker}.

This null result joins the ranks of past dark matter annihilation 
searches, which have failed
to detect a gamma-ray signal in systems
suspected of high dark matter content \citep{bringmann}.
Unfortunately, as with the rest of dark matter pursuits, 
this is an everything or nothing undertaking. 
Here, we are further hampered by the fact that we are seeking a detection 
with nearly no information about distances and masses of the objects
involved. 
Nonetheless, {\it Fermi} will continue to collect data and stricter gamma-ray 
limits for these systems 
can be reached. Above $E \gtrsim $ 100 GeV, the Cherenkov Telescope
Array (CTA) will be crucial to escalate the 
dark matter search to unprecedented bounds \citep{cta,doro}.
There is also sufficient motivation  
to explore signatures for other reasonable dark matter 
candidates at other wavelengths \citep{feng}.

From our measurements, it 
is still unclear where GALFA-HI galaxy candidates fit
in the larger dark matter subhalo picture. 
Searches for stellar counterparts associated with
these gas-bearing systems are underway and should 
intensify \citep{saul}. If this goal is accomplished, 
member stars could be used to produce a dark matter density profile. 
In an optimistic scenario, 
compact clouds could validate a sort of ``hiding in plain sight'' model
whereby 
dark matter subhalos at small scales would be traced directly through 
neutral hydrogen.

Alternatively, dedicated observational studies could finally 
establish that all compact 
clouds are purely baryonic and hence 
devoid of dark matter \citep{plockinger}. The disagreement between 
observations and $\Lambda$CDM simulations 
might then have to be invoke more inventive solutions \citep{boylan}.
Regardless, the potential dark matter 
fingerprint discussed here stands a possible diagnostic of 
suspected nearby dark matter dominated galactic candidates.  
With the advent of Skymapper \citep{keller} and the Large
Synoptic Survey Telescope \citep{ivezic}, it might be possible to
trace the subhalo population directly using faint stars. 
A cross-match between stellar concentrations and
compact hydrogen cloud positions should
be conducted as soon as said surveys are completed.

\section*{Acknowledgments}
We thank the referee for useful suggestions that improved the paper. 
N.M. acknowledges support from the Spanish government 
through a Ram\'on y Cajal fellowship, the 
Consolider-Ingenio 2010 Programme under grant MultiDark CSD2009-00064,
and the Spanish MINECO under project code 
FPA2010-22056-C06-06. 
We also acknowledge the use of public data from the {\it Fermi} 
data archives. 
Lastly, we thank Jeremy S. Perkins for contributing a 
wonderful set of Python libraries  for routine {\it Fermi} analysis.

\label{lastpage}
\end{document}